\documentclass{proc}
\usepackage{amsfonts}
\usepackage{amsmath}
\usepackage{graphicx}

\catcode`\@=11

\let\DOTSI\relax
\def\RIfM@{\relax\ifmmode}
\def\FN@{\futurelet\next}
\newcount\intno@
\def\iint{\DOTSI\intno@\tw@\FN@\ints@}
\def\iiint{\DOTSI\intno@\thr@@\FN@\ints@}
\def\iiiint{\DOTSI\intno@4 \FN@\ints@}
\def\idotsint{\DOTSI\intno@\z@\FN@\ints@}
\def\ints@{\findlimits@\ints@@}
\newif\iflimtoken@
\newif\iflimits@
\def\findlimits@{\limtoken@true\ifx\next\limits\limits@true
 \else\ifx\next\nolimits\limits@false\else
 \limtoken@false\ifx\ilimits@\nolimits\limits@false\else
 \ifinner\limits@false\else\limits@true\fi\fi\fi\fi}
\def\multint@{\int\ifnum\intno@=\z@\intdots@                                
 \else\intkern@\fi                                                          
 \ifnum\intno@>\tw@\int\intkern@\fi                                         
 \ifnum\intno@>\thr@@\int\intkern@\fi                                       
 \int}                                                                      
\def\multintlimits@{\intop\ifnum\intno@=\z@\intdots@\else\intkern@\fi
 \ifnum\intno@>\tw@\intop\intkern@\fi
 \ifnum\intno@>\thr@@\intop\intkern@\fi\intop}
\def\intic@{\mathchoice{\hskip.5em}{\hskip.4em}{\hskip.4em}{\hskip.4em}}
\def\negintic@{\mathchoice
 {\hskip-.5em}{\hskip-.4em}{\hskip-.4em}{\hskip-.4em}}
\def\ints@@{\iflimtoken@                                                    
 \def\ints@@@{\iflimits@\negintic@\mathop{\intic@\multintlimits@}\limits    
  \else\multint@\nolimits\fi                                                
  \eat@}                                                                    
 \else                                                                      
 \def\ints@@@{\iflimits@\negintic@
  \mathop{\intic@\multintlimits@}\limits\else
  \multint@\nolimits\fi}\fi\ints@@@}
\def\intkern@{\mathchoice{\!\!\!}{\!\!}{\!\!}{\!\!}}
\def\plaincdots@{\mathinner{\cdotp\cdotp\cdotp}}
\def\intdots@{\mathchoice{\plaincdots@}
 {{\cdotp}\mkern1.5mu{\cdotp}\mkern1.5mu{\cdotp}}
 {{\cdotp}\mkern1mu{\cdotp}\mkern1mu{\cdotp}}
 {{\cdotp}\mkern1mu{\cdotp}\mkern1mu{\cdotp}}}

\newif\iffirstchoice@
\firstchoice@true
\def\textfonti{\the\textfont\@ne}
\def\textfontii{\the\textfont\tw@}
\def\text{\RIfM@\expandafter\text@\else\expandafter\text@@\fi}
\def\text@@#1{\leavevmode\hbox{#1}}
\def\text@#1{\mathchoice
 {\hbox{\everymath{\displaystyle}\def\textfonti{\the\textfont\@ne}%
  \def\textfontii{\the\textfont\tw@}\textdef@@ T#1}}
 {\hbox{\firstchoice@false
  \everymath{\textstyle}\def\textfonti{\the\textfont\@ne}%
  \def\textfontii{\the\textfont\tw@}\textdef@@ T#1}}
 {\hbox{\firstchoice@false
  \everymath{\scriptstyle}\def\textfonti{\the\scriptfont\@ne}%
  \def\textfontii{\the\scriptfont\tw@}\textdef@@ S\rm#1}}
 {\hbox{\firstchoice@false
  \everymath{\scriptscriptstyle}\def\textfonti
  {\the\scriptscriptfont\@ne}%
  \def\textfontii{\the\scriptscriptfont\tw@}\textdef@@ s\rm#1}}}
\def\textdef@@#1{\textdef@#1\rm\textdef@#1\bf\textdef@#1\sl\textdef@#1\it}
\def\DN@{\def\next@}
\def\eat@#1{}
\def\textdef@#1#2{%
 \DN@{\csname\expandafter\eat@\string#2fam\endcsname}%
 \if S#1\edef#2{\the\scriptfont\next@\relax}%
 \else\if s#1\edef#2{\the\scriptscriptfont\next@\relax}%
 \else\edef#2{\the\textfont\next@\relax}\fi\fi}

\def\Let@{\relax\iffalse{\fi\let\\=\cr\iffalse}\fi}
\def\vspace@{\def\vspace##1{\crcr\noalign{\vskip##1\relax}}}
\def\multilimits@{\bgroup\vspace@\Let@
 \baselineskip\fontdimen10 \scriptfont\tw@
 \advance\baselineskip\fontdimen12 \scriptfont\tw@
 \lineskip\thr@@\fontdimen8 \scriptfont\thr@@
 \lineskiplimit\lineskip
 \vbox\bgroup\ialign\bgroup\hfil$\m@th\scriptstyle{##}$\hfil\crcr}
\def\Sb{_\multilimits@}
\def\endSb{\crcr\egroup\egroup\egroup}
\def\Sp{^\multilimits@}

\newdimen\ex@
\def\rightarrowfill@#1{$#1\m@th\mathord-\mkern-6mu\cleaders
 \hbox{$#1\mkern-2mu\mathord-\mkern-2mu$}\hfill
 \mkern-6mu\mathord\rightarrow$}
\def\leftarrowfill@#1{$#1\m@th\mathord\leftarrow\mkern-6mu\cleaders
 \hbox{$#1\mkern-2mu\mathord-\mkern-2mu$}\hfill\mkern-6mu\mathord-$}
\def\leftrightarrowfill@#1{$#1\m@th\mathord\leftarrow\mkern-6mu\cleaders
 \hbox{$#1\mkern-2mu\mathord-\mkern-2mu$}\hfill
 \mkern-6mu\mathord\rightarrow$}
\def\overrightarrow{\mathpalette\overrightarrow@}
\def\overrightarrow@#1#2{\vbox{\ialign{##\crcr\rightarrowfill@#1\crcr
 \noalign{\kern-\ex@\nointerlineskip}$\m@th\hfil#1#2\hfil$\crcr}}}

\def\overleftarrow{\mathpalette\overleftarrow@}
\def\overleftarrow@#1#2{\vbox{\ialign{##\crcr\leftarrowfill@#1\crcr
 \noalign{\kern-\ex@\nointerlineskip}$\m@th\hfil#1#2\hfil$\crcr}}}
\def\overleftrightarrow{\mathpalette\overleftrightarrow@}
\def\overleftrightarrow@#1#2{\vbox{\ialign{##\crcr\leftrightarrowfill@#1\crcr
 \noalign{\kern-\ex@\nointerlineskip}$\m@th\hfil#1#2\hfil$\crcr}}}
\def\underrightarrow{\mathpalette\underrightarrow@}
\def\underrightarrow@#1#2{\vtop{\ialign{##\crcr$\m@th\hfil#1#2\hfil$\crcr
 \noalign{\nointerlineskip}\rightarrowfill@#1\crcr}}}

\def\underleftarrow{\mathpalette\underleftarrow@}
\def\underleftarrow@#1#2{\vtop{\ialign{##\crcr$\m@th\hfil#1#2\hfil$\crcr
 \noalign{\nointerlineskip}\leftarrowfill@#1\crcr}}}
\def\underleftrightarrow{\mathpalette\underleftrightarrow@}
\def\underleftrightarrow@#1#2{\vtop{\ialign{##\crcr$\m@th\hfil#1#2\hfil$\crcr
 \noalign{\nointerlineskip}\leftrightarrowfill@#1\crcr}}}

\catcode`\@=\active

\def\frac#1#2{{#1 \over #2}}

\newcount\GRAPHICSTYPE
\def\GRAPHICSPS#1{%
\ifnum\GRAPHICSTYPE=1 language "PS", include "#1"\else%
ps: #1\fi}

\def\graffile#1#2#3#4{\leavevmode\raise -#4 \hbox{%
\raise #3 \hbox{\rule{0.003in}{0.003in}\special{#1}}}%
{\raise -#4 \hbox to #2 {\vrule height#3 width0in depth0in\hfil}}%
}

\def\draftbox#1#2#3#4{\leavevmode\raise -#4 \hbox{\frame{\rlap{\protect\tiny #1}%
\hbox to #2{\vrule height#3 width0in depth0in\hfil}}}}

\newcount\draft
\def\GRAPHIC#1#2#3#4#5{\ifnum\draft=1 \draftbox{#2}{#3}{#4}{#5}\else%
\graffile{#1}{#3}{#4}{#5}\fi}

\def\addtoLaTeXparams#1{\edef\LaTeXparams{\LaTeXparams #1}}

\def\doFRAMEparams#1{\readFRAMEparams#1\end}
\def\readFRAMEparams#1{%
\ifx#1\end%
\let\next=\relax%
\else%
\ifx#1i%
\dispkind=0%
\fi%
\ifx#1d%
\dispkind=1%
\fi%
\ifx#1f%
\dispkind=2%
\fi%
\ifx#1t%
\addtoLaTeXparams{t}%
\fi%
\ifx#1b%
\addtoLaTeXparams{b}%
\fi%
\ifx#1p%
\addtoLaTeXparams{p}%
\fi%
\ifx#1h%
\addtoLaTeXparams{h}%
\fi%
\let\next=\readFRAMEparams%
\fi%
\next%
}

\def\IFRAME#1#2#3#4#5{\GRAPHIC{#5}{#4}{#1}{#2}{#3}}

\def\DFRAME#1#2#3#4{
  \begin{center}
    \GRAPHIC{#4}{#3}{#1}{#2}{0in} 
  \end{center}
}

\def\FFRAME#1#2#3#4#5#6#7{
  \begin{figure}[#1]
    \begin{center}
      \GRAPHIC{#7}{#6}{#2}{#3}{0in}
    \end{center}
    \caption{\label{#5}#4}
  \end{figure}
}

\def\FRAME#1#2#3#4#5#6#7#8{%
\def\LaTeXparams{}%
\dispkind=0%
\def\LaTeXparams{}%
\doFRAMEparams{#1}%
\ifnum\dispkind=0%
\IFRAME{#2}{#3}{#4}{#7}{#8}%
\else
  \ifnum\dispkind=1
    \DFRAME{#2}{#3}{#7}{#8}
  \else
    \ifnum\dispkind=2
      \FFRAME{\LaTeXparams}{#2}{#3}{#5}{#6}{#7}{#8}
    \fi
  \fi
\fi
}

\catcode`\@=11

\long\def\QQQ#1#2{}
\def\QTP#1{}
\long\def\QQA#1#2{}

\def\EXPAND#1[#2]#3{}
\def\NOEXPAND#1[#2]#3{}

\def\LaTeXparent#1{}

\def\initial#1{\bigbreak{\raggedright\large\bf #1}\kern 2pt\penalty3000}

\newdimen\theight
\def \Column{%
             \vadjust{\setbox0=\hbox{\scriptsize\quad\quad tcol}%
             \theight=\ht0
             \advance\theight by \dp0    \advance\theight by \lineskip
             \kern -\theight \vbox to \theight{\rightline{\rlap{\box0}}%
             \vss}%
             }}%

\def\qed{\ifhmode\unskip\nobreak\fi\ifmmode\ifinner\else\hskip5\p@\fi\fi
 \hbox{\hskip5\p@\vrule width4\p@ height6\p@ depth1.5\p@\hskip\p@}}
\catcode`@=12 

\begin{document}

\title{Mining Urban Performance: Scale-Independent Classification of Cities Based on Individual Economic Transactions}
\author{S. Sobolevsky$^{1}$ , I. Sitko$^{2}$, S. Grauwin$^{1}$, R. Tachet des Combes$^{1}$, B. Hawelka$^{2}$, J.Murillo Arias$^{3}$, C. Ratti$^{1}$ \\
$^{1}$SENSEeable City Lab, MIT\\
$^{2}$Department of Geoinformatics - Z\_GIS, University of Salzburg\\
$^{3}$New Technologies, BBVA\\
stanly@mit.edu, izabela.sitko@sbg.ac.at, sgrauwin@mit.edu, rtachet@mit.edu, \\bartosz.hawelka@sbg.ac.at, juan.murillo.arias@bbva.com, ratti@mit.edu}
\maketitle

\begin{abstract}
Intensive development of urban systems creates a number of challenges for urban planners and policy makers in order to maintain sustainable growth. Running efficient urban policies requires meaningful urban metrics, which could quantify important urban characteristics including various aspects of an actual human behavior. Since a city size is known to have a major, yet often nonlinear, impact on the human activity, it also becomes important to develop scale-free metrics that capture qualitative city properties, beyond the effects of scale. Recent availability of extensive datasets created by human activity involving digital technologies creates new opportunities in this area. In this paper we propose a novel approach of city scoring and classification based on quantitative scale-free metrics related to economic activity of city residents, as well as domestic and foreign visitors. It is demonstrated on the example of Spain, but the proposed methodology is of a general character. We employ a new source of large-scale ubiquitous data, which consists of anonymized countrywide records of bank card transactions collected by one of the largest Spanish banks. Different aspects of the classification reveal important properties of Spanish cities, which significantly complement the pattern that might be discovered with the official socioeconomic statistics.
\end{abstract}

\section{Introduction}

In the rapidly urbanizing world, cities have become main centers of human activity, gathering over half of the global population \cite{un2011}. Intensive development of urban systems creates many opportunities, but also challenges for urban planners and policy makers in order to maintain sustainable growth. Adequate and reliable metrics of city performance lay in the core of successful management. In the context of an increasingly competitive economy, prosperity of a city needs to be judged not only in absolute value, but also in relation to its theoretical potential and to the performance of surrounding urban centers \cite{arribas-bel2013}. In practice, this need is often addressed through various types of city rankings and classifications, e.g. \cite{giffinger2007, caragliu2012, citiesoutlook2012}. However, direct juxtaposition of cities' performance raises at least two questions, connected to: (i) the availability and exploitation of a uniform and up-to-date dataset (ii) the varying character and size of cities to be compared. In this paper, we try to tackle both issues using a new type of pervasive data, individual bank cards transactions made in a city and by its citizens, to construct scale-free metrics of city performance.

Important challenge in city comparison come from the data side. Because they are provided for an arbitrary set of administrative units and preselected time stamps, traditional city statistics and indicators are often static and of a selective character. The overwhelming digitalization of everyday life has recently opened up new possibilities for human behavioral studies, also in urban environments. The digital footprints of human activity, such as mobile phone records, social media or smart cards usage, have already demonstrated substantial potential to address important problems of an economic geography, e.g. regional delineation \cite{blondel2010regions, ratti2010redrawing, sobolevsky2013delineating, hawelka2014}, they have also been intensively investigated as city descriptors \cite{reades2007, girardin2009, cranshaw2012, lathia2012, kang2013exploring, Amini2014}. Thorough analysis of individual bank card records has been quite limited so far, a fact most probably due to obvious access constraints. Hitherto, applications were mostly focused on issues within the card system itself \cite{chan1999frauddetection, rysman2007}, rather than on the associated human behavior. The only exception known to the authors is the study of Krumme et al. \cite{krumme2010patterns}, who examined predictability of human spending based on their loyalty towards merchants types. With this exception in mind, bank card data has not been yet explored in the urban context, a gap we are willing to fill here.

Given a consistent dataset, the second factor hindering comparisons between cities is their varying size (generally defined as their population) - from several thousands to several millions of inhabitants. This inequality is conventionally dealt with {\it per capita} metrics, assuming a linear relationship between size and city parameters. Recently however, substantial research points towards a non-linear dependence on city size \cite{bettencourt2007growth, batty2008size}. Due to agglomeration effects and intensified human interactions, a variety of city metrics were empirically observed to change systematically with city population in the form of scaling laws \cite{bettencourt2013origins, bettencourt2007growth, brockmann2006scaling, schlapfer2012scaling}. While socioeconomic quantities reveal a superlinear relation to city size, urban infrastructure dimensions, e.g. total road surface, usually decrease in a sublinear manner \cite{bettencourt2013origins}.

In this paper, we focused on three urban parameters extracted from bank card transactions: total spending of a given city inhabitants, total spending in a given city by a country residents and total spending in a given city by foreign residents. In order to built a consistent metric of a city's performance on those three values, we started by studying their dependence on city size. In a first section, the impact of total population on spending activity and attractiveness for potential visitors is established, and quantified by a superlinear scaling. This allowed to relate the performance of each city to its estimated value, computed by means of the aforementioned law. Observed deviations indicated the actual performance of the city, indirect effects of its size taken into account. This approach enabled a consistent comparison of urban performance between cities of different sizes. Following the general approach of \cite{bettencourt2010urbscaling} we used appropriately quantified deviations from the general scaling trends as the individual scale-independent city performance index. Finally, we shown how classifying the cities using those residual metrics can reveal important geographical, social, economic and cultural patterns.

\section{The Dataset}

Our study relies on the complete set of bank card transactions recorded by Banco Bilbao Vizcaya Argentaria (BBVA) during 2011, all over Spain\footnote{The raw dataset is protected by the appropriate non-disclosure agreement and is not publicly available. However, the researchers may share certain aggregated data upon request and for the purpose of findings validation.}. Those transactions were performed by two groups of card users. The first one consists of the bank direct customers, residents of Spain, who hold a debit or credit card issued by BBVA. In 2011, the total number of active customers was around 4.5\,M, altogether they executed more than 178\,M transactions in over 1.2\,M points of sale, spending over 10 billion euros. The second group of card users includes over 8.6\,M foreign customers of all other banks abroad coming from 175 countries\footnote{There are countries that are not represented in our dataset due to the legal and/or technical restrictions of bank data access, the most significant example being Germany. Also, certain countries do not include significant number of people who visit Spain and use their national bank cards.}, who made purchases through one of the approximately 300 thousand BBVA card terminals. In total, they executed another 17M transactions, spending over 1.5\,B euro.

Due to the sensitive nature of bank data, our dataset was anonymized by BBVA prior to sharing, in accordance to all local privacy protection laws and regulations. As a result, customers are identified by randomly generated IDs, connected with certain demographic characteristics and an indication of a residence location - at the level of zip code for direct customers of BBVA and country of residence for all others. Each transaction is denoted with its value, a time stamp, a retail location where it was performed, and the business category it belongs to. The business classification includes 76 categories, which were further aggregated into 12 meaningful major groups, gathering e.g. purchases of food, fashion, home appliances or travel activities.

Before the analysis, the dataset had to be pre-processed in order to compensate for the potential bias introduced by the inhomogeneity of BBVA market share. First type of bias concerned representativeness of BBVA customers as the varying fraction of an economically active population, depending on the BBVA penetration of the individual banking market in a given area. Therefore, customers' activity was normalized by the respective market share in their residence location, provided by the bank at the level of province, in order to estimate total domestic customer spending. Another type of bias concerned foreign customers executing transactions in the BBVA point of sale terminals, unequally distributed inside different cities and across the regions of Spain.  In this case the normalization procedure relied on the BBVA business market share, defined for the purpose of this study as the ratio between the transactions executed by domestic customers in the BBVA terminals to their transactions in all other terminals located in a considered area. Appropriate normalization allowed to estimate total spending of foreign customers visiting particular city.

Comparison across different cities requires clear and consistent definition of their spatial extent. The ideal units should reflect functional character of cities and social interactions of their inhabitants, rather than purely administrative division \cite {bettencourt2010urbscaling, betten}. In our study we tested two levels of city definition taking advantage of units provided by European Urban Audit Survey \cite{urbauditweb}. The higher level consisted of 24  Large Urban Zones (LUZs) which reflect larger urban agglomerations \cite{urbaudit2004}. For the finer level we aggregated Administrative Cities of Urban Audit \cite{urbaudit2004} into 40 Functional Urban Areas (FUAs), so as to reflect metropolitan regions in agreement with the Study on Urban Functions of the European Spatial Planning Observation Network (ESPON) \cite{espon2007}. Population and economic statistics for both city levels were obtained from Eurostat \cite{eurostat} and National Statistics Institute of Spain \cite{ine}.

\section{Major urban parameters and their scaling with city size}

With the evergrowing urbanization, cities are the focus of tremendous research efforts, to understand in particular what drives us to live in such entities and what we gain from them. Negatively considered at the beginning of the twentieth century \cite{simmel, wirth}, it has been shown since then, and with great accuracy in a recent past that cities are actually the vector of a more intense activity. Wages, gross product, patents, creativity, as well as crimes and diseases behave in a superlinear way with the urban area population \cite{bettencourt2007invention, bettencourt2007growth, batty2008size, bettencourt2010urbscaling}.

In the present research, we started looking for similar behaviors in bank card transactions. The dataset at our disposal allowed us to describe economic performance of a city from several compatible perspectives. For a given town, the activity levels of the residents, combined with various characteristics of said activity can be seen as a measure of its internal prosperity and purchasing potential. Activity of domestic customers including those coming from other locations in Spain indicates the prosperity of the city businesses and also what kind of part the city plays on the regional or perhaps national stage, attracting people and resources. Eventually, and on a larger scale, activity of foreign visitors unveils the city attractiveness as an international touristic or business destination. In accordance with those statements, we thus defined city performance on the corresponding three parameters: 1) total activity of city residents (referred to as resident activity or RA), quantified as the total number of transactions executed countrywide by inhabitants of the city; 2) total activity recorded in retail points belonging to the city and executed by Spanish customers (city domestic earnings or CDE in the following) defined as the total number of transactions made in the city by domestic customers coming from anywhere in Spain; 3) city attractiveness for foreign visitors (referred to as city foreign attractiveness or FA) equal to the total number of transactions executed in the city by foreign customers. 

\begin{figure*}[ht!]
\centering
\includegraphics[width=0.45\linewidth]{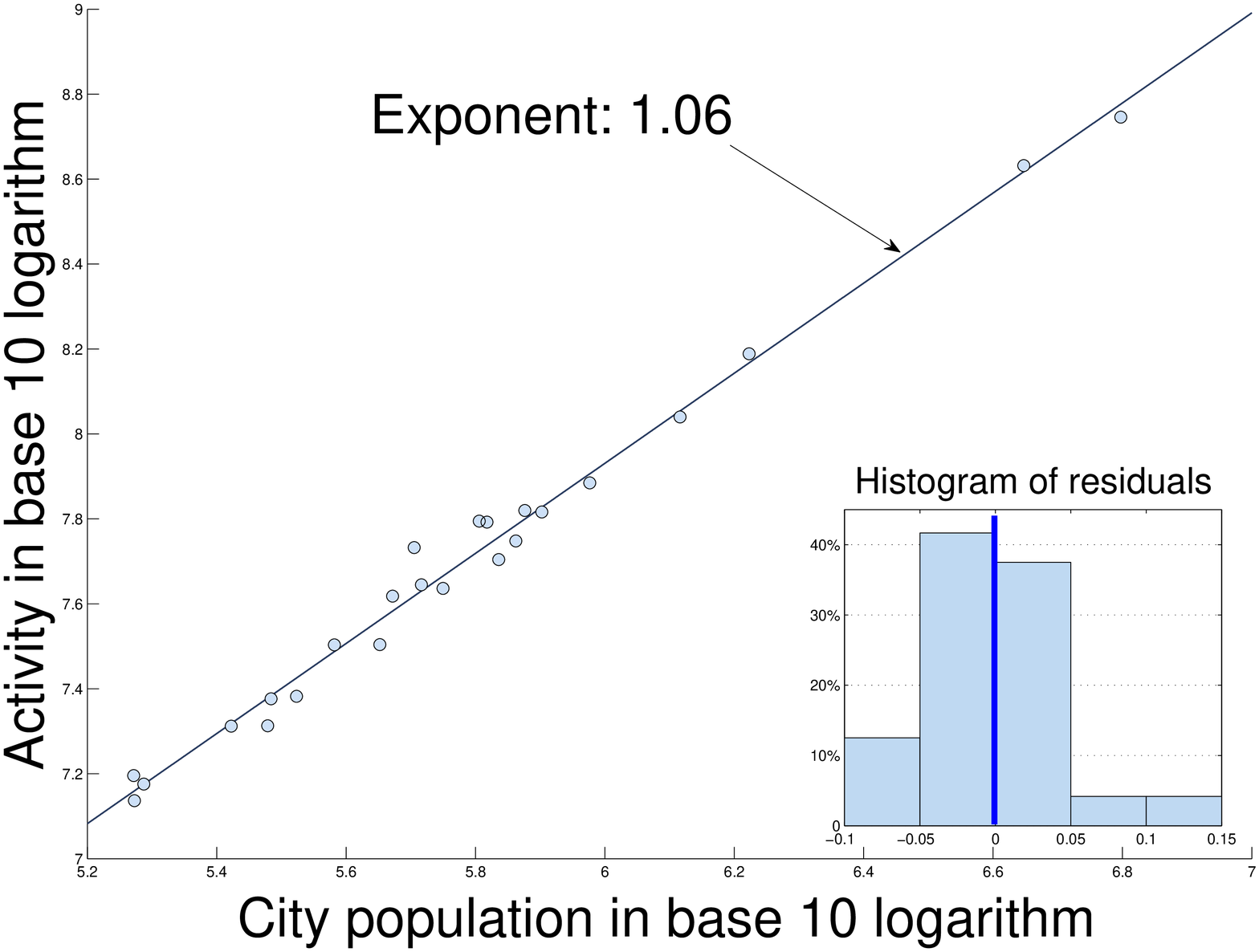}
\includegraphics[width=0.45\linewidth]{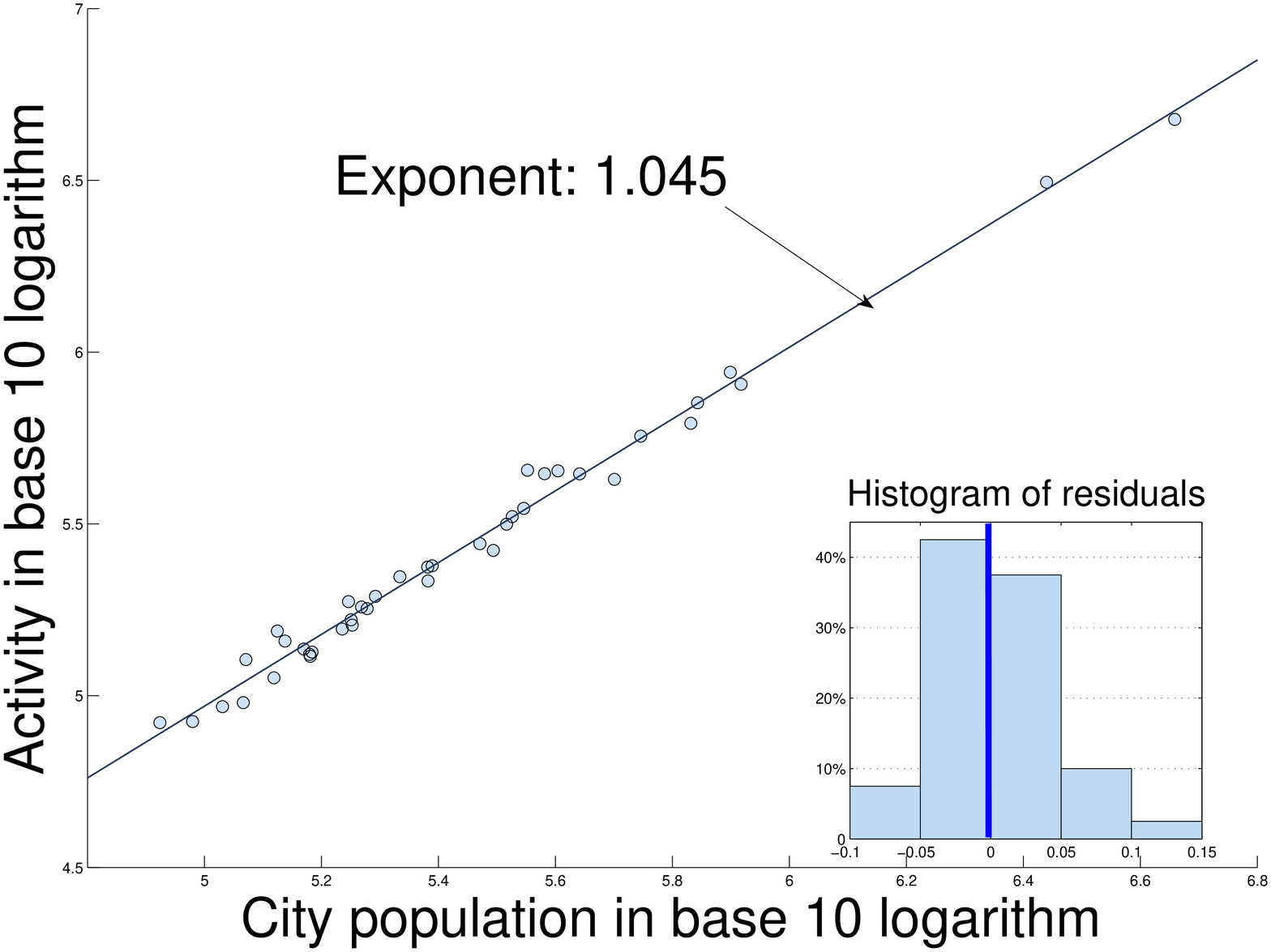}
\includegraphics[width=0.45\linewidth]{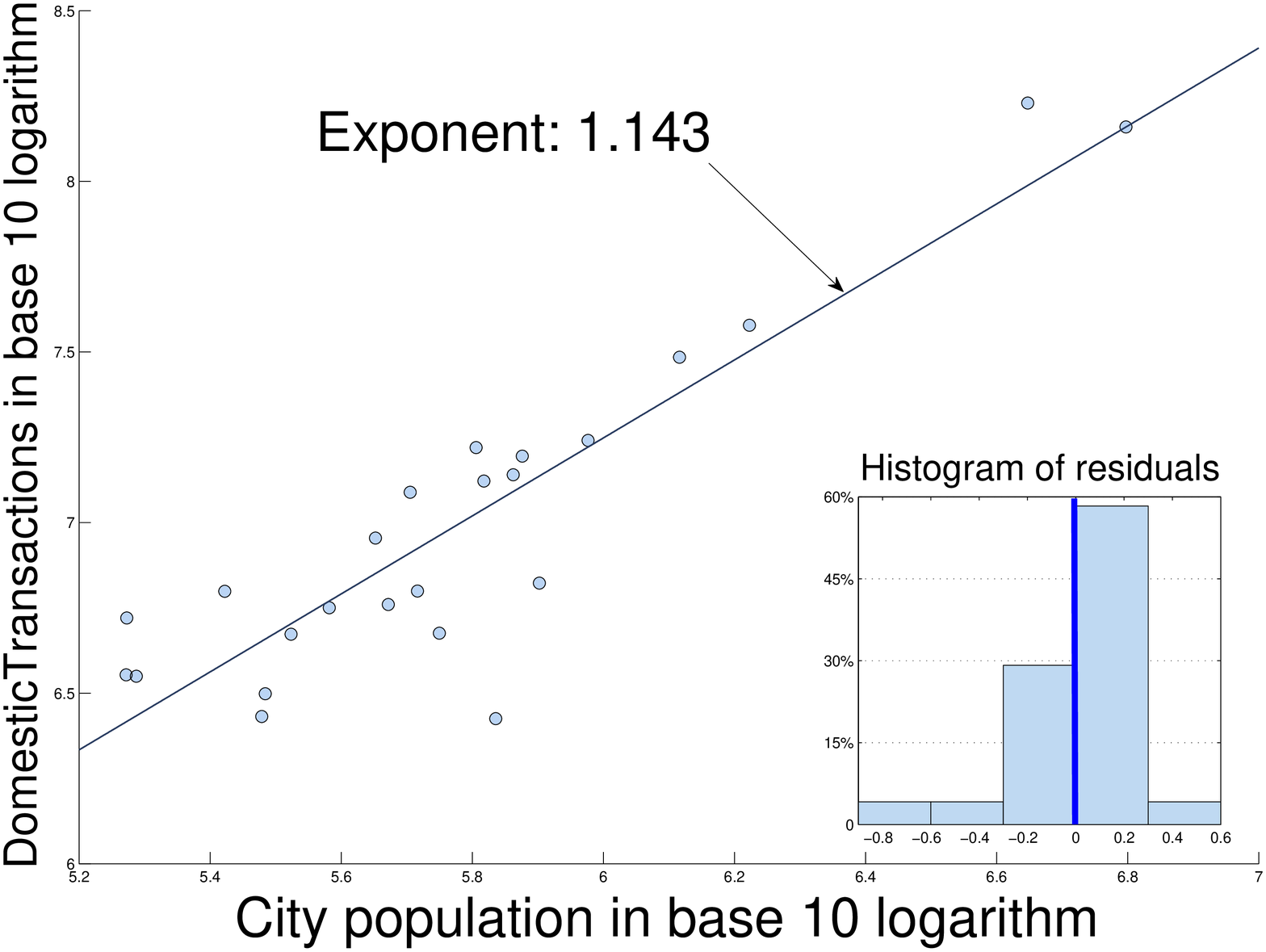}
\includegraphics[width=0.45\linewidth]{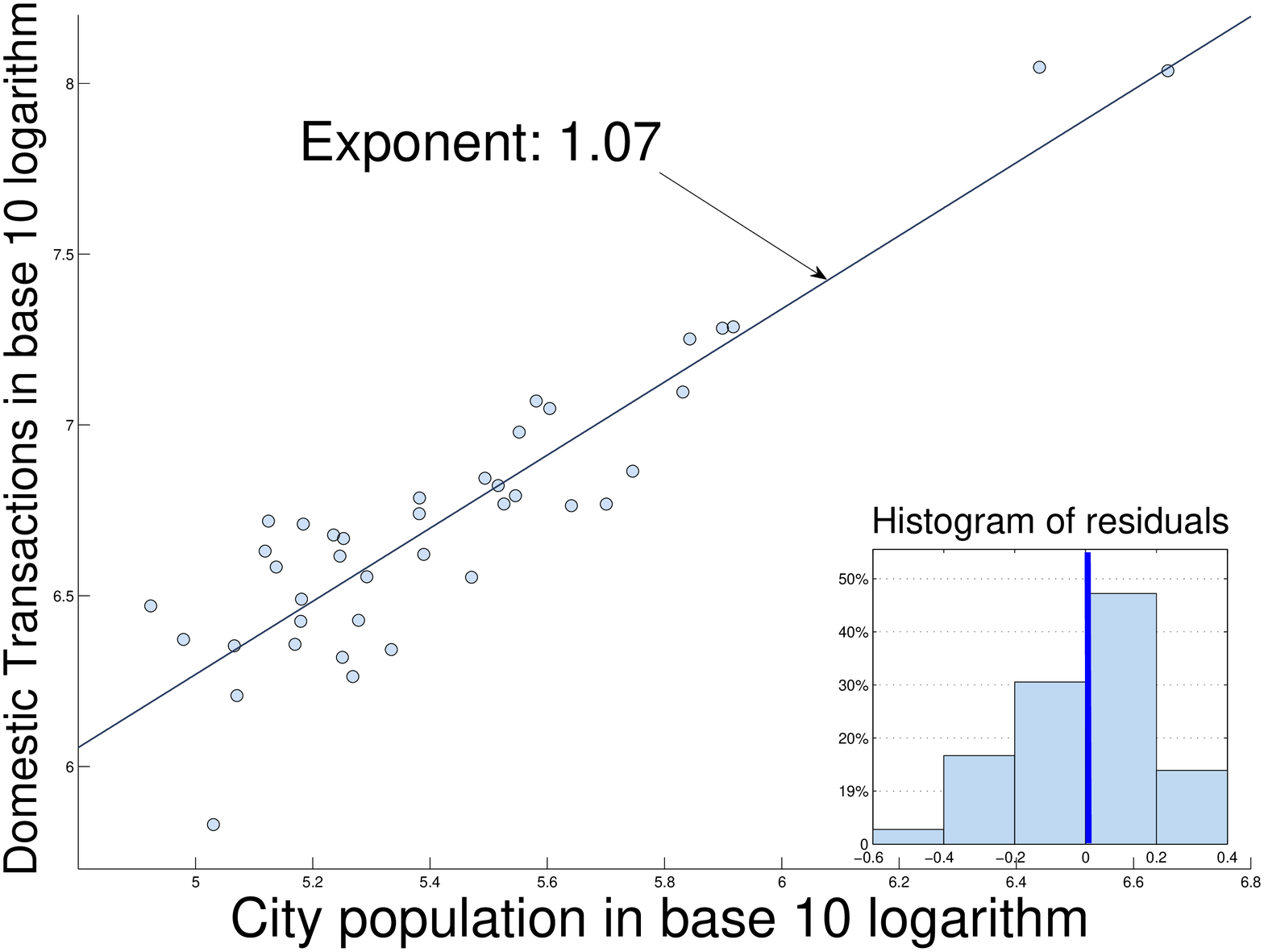}
\includegraphics[width=0.45\linewidth]{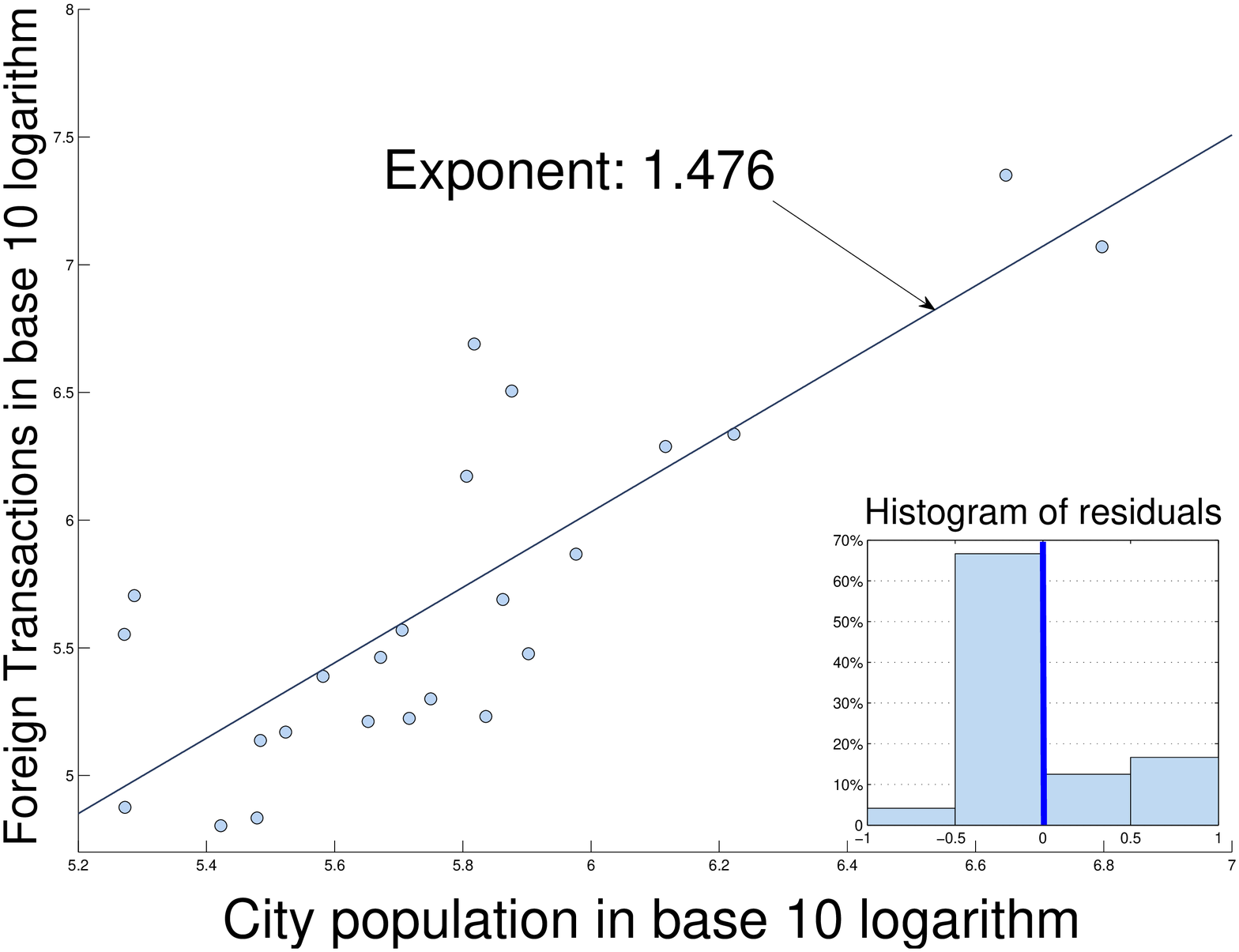}
\includegraphics[width=0.45\linewidth]{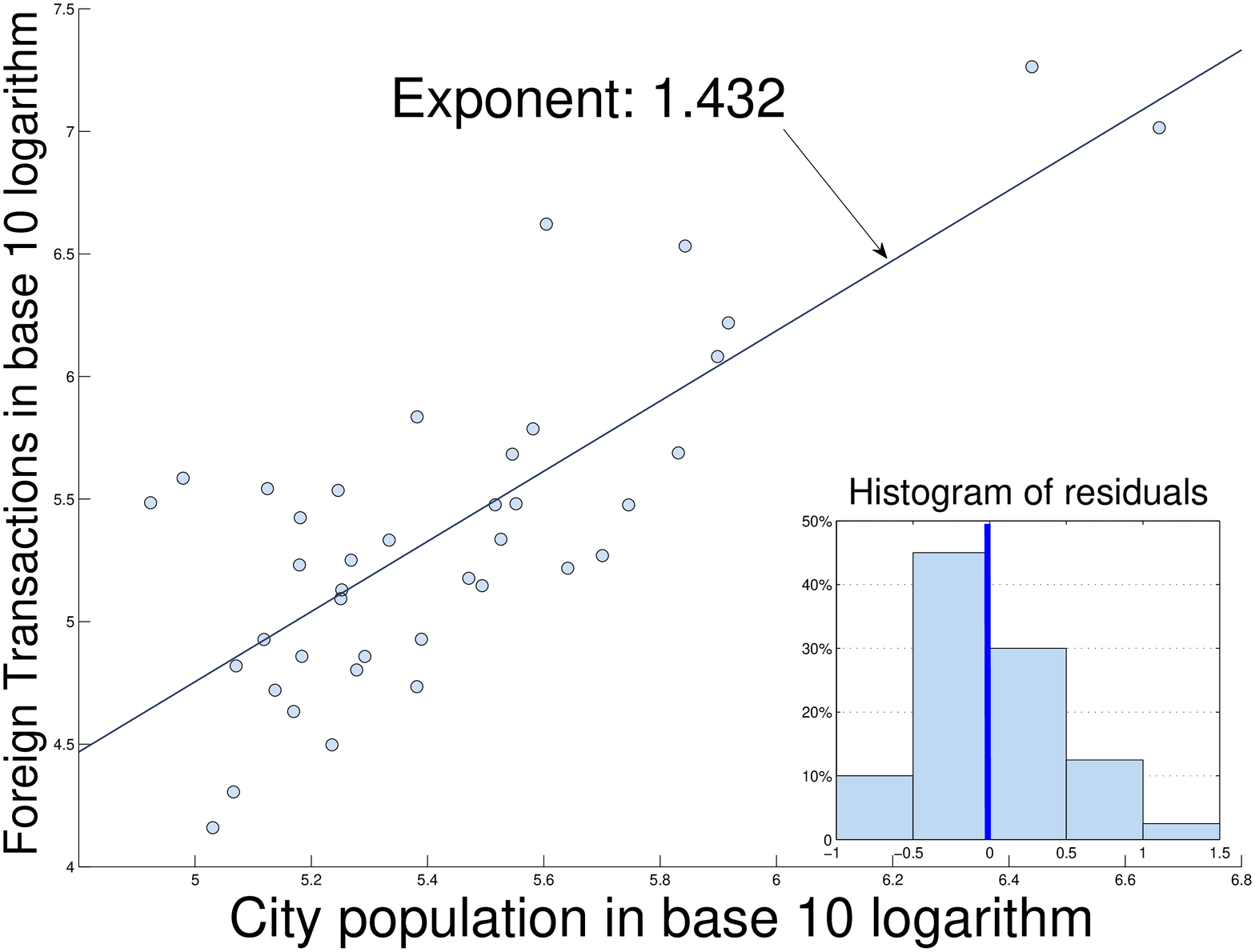}
\caption{\label{fig::scaling}Superlinear scaling of major city parameters with city size - residential activity, domestic earnings and foreign attractiveness from top to bottom. Large Urban Zones on the left, Functional Urban Areas on the right. Subplots present the distribution of log residuals from the respective scaling trend.}
\end{figure*}

In \cite{bettencourt2007growth}, the pace of urban life is shown to behave as a power-law of city size with exponents larger than 1. Transactional activity measures such a pace from a different angle, we thus study it to see whether it varies in a similar way. The log-log plot of the aforementioned parameters against population can be found in figure \ref{fig::scaling}, separately for the two considered levels of a city definition: 24 Large Urban Zones (LUZ) and 40 Functional Urban Areas (FUA). As expected, all parameters demonstrate a superlinear scaling. The scaling exponents are generally consistent for both city levels, with the only exception being CDE. Especially strong is the superlinearity of the FA trend - the exponent of nearly $1.5$ evidences that if one city is e.g. 3 times bigger than another, its attractiveness is around 5 times higher on average. It points out to a strong pattern: foreign visitors are predominantly attracted to the largest destinations in the country.

\section{Residuals from the scaling trends as scale-independent metrics of a~city performance}
 
General scaling laws describe how urban parameters are expected to change with city size \cite {bettencourt2007growth, batty2008size}. In reality, actual values of a parameter always deviate from a theoretical trend due to the additional influence of local city characteristics. A value that is higher than a trend estimate can be treated as the indication of a city over-performance in a given domain. On the contrary, values below a scaling trend may be interpreted as the under-performance. Deviations of different cities are assessed in relation to the size-specific estimates, which makes them directly comparable, beyond the impact of scale. Therefore, this kind of deviations can be used as a scale-independent metric of a qualitative city performance. Similarly to \cite{bettencourt2010urbscaling}, we quantified those deviations as the log-scale residuals, i.e., the decimal logarithm of an actual city characteristic minus the decimal logarithm of the corresponding trend estimation for a given city size. 

\begin{figure*}[ht!]
\centering
\vspace{5pt}
\includegraphics[width=0.95\linewidth]{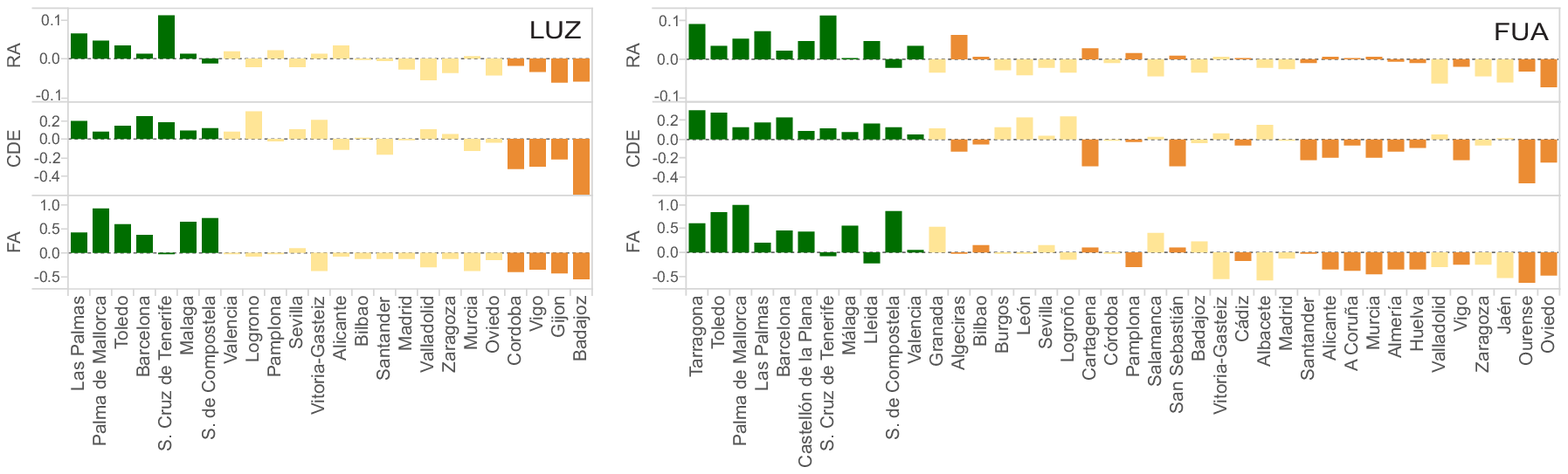}
\vspace{-5pt}
\caption{\label{fig::residuals}City performance metrics quantified as the log residuals of RA, CDE and FA parameters from their scaling trends with city size. Colors indicate three categories of cities obtained based on the clustering of the residuals (for spatial distribution of the clusters see figure  \ref{fig::clustering3}).Cities are ordered according to their total performance, defined with the sum of normalized ranks in all three parameters.}
\end{figure*}

Residuals obtained for each of the examined urban parameters, RA, CDE and FA, and for both levels of city definition, are presented in figure  \ref{fig::residuals}. We can observe different types of city profiles. Some of them, e.g. Toledo, Barcelona or Palma de Mallorca, record positive values across all three metrics, while others, such as Vigo or Badajoz,  show a clear tendency to under-perform regardless of the parameter. Finally, there is a significant group of cities with mixed scores - positive for some and negative for other metrics. This variability is visibly higher at the finer city level, however, profiles of the corresponding cities show general consistency between the levels of LUZ and FUA. 

\begin{figure*}[ht!]
\centering
\vspace{-5pt}
\includegraphics[width=0.43\linewidth]{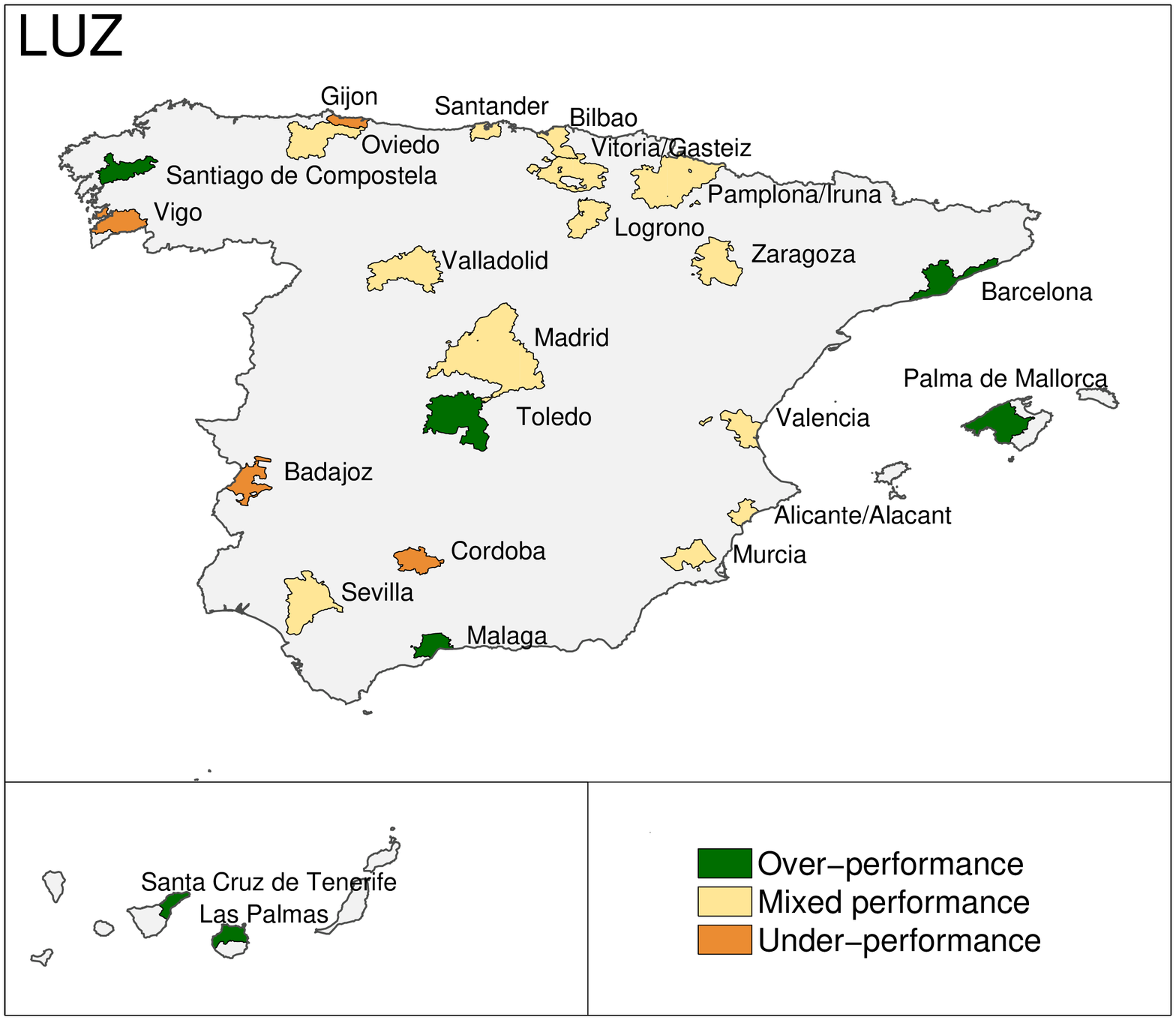}
\includegraphics[width=0.43\linewidth]{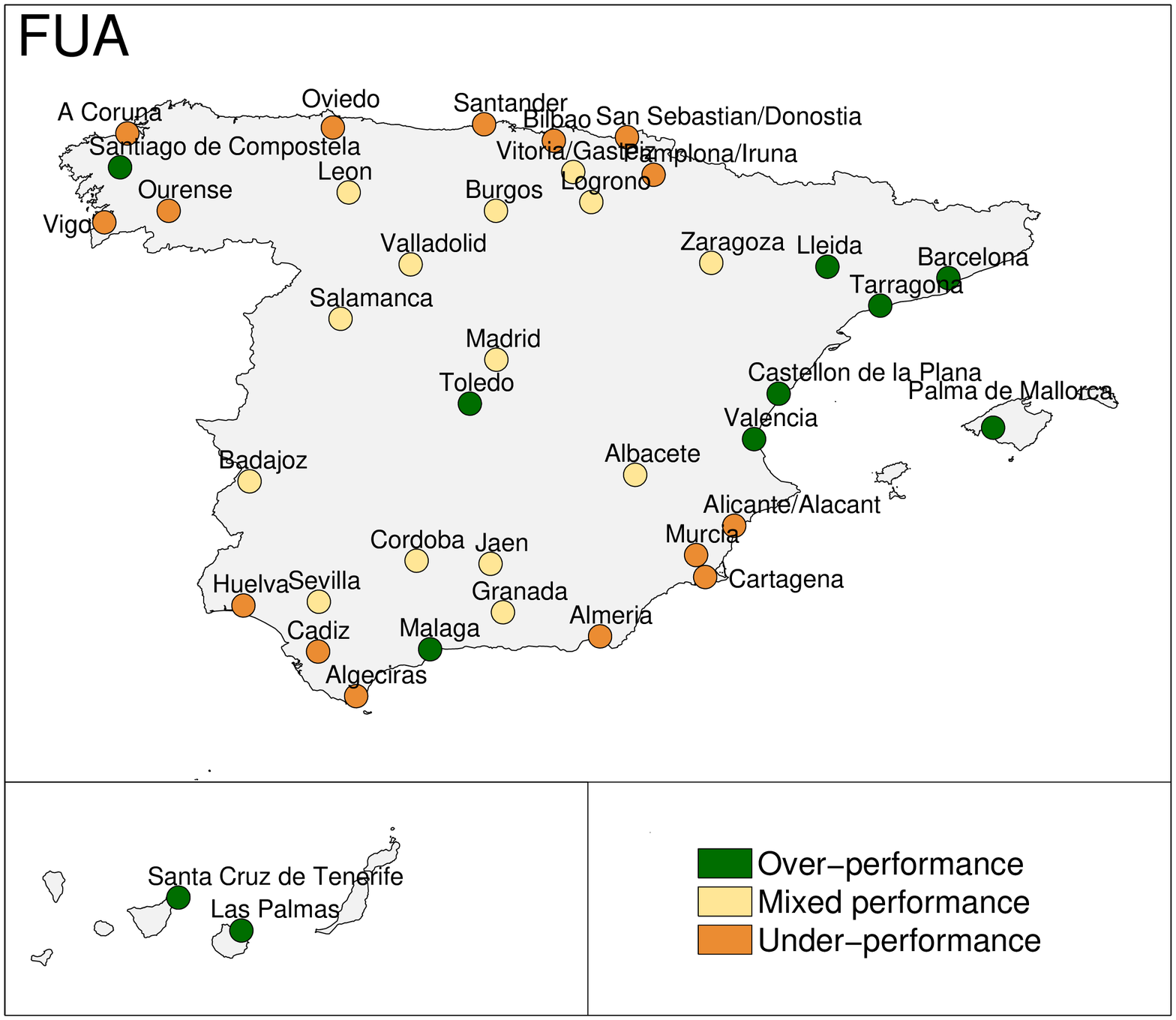}
\vspace{-20pt}
\caption{\label {fig::clustering3}Different categories of cities obtained based on the clustering of scale-independent city performance metrics into k=2,3 clusters. The partitioning for k=2 can be obtained by considering yellow and orange clusters (mixed and under-performance) as one - with the only exception of Granda at the level of FUA, which is assigned to the green cluster (over-performance) for the case of k=2.}
\end{figure*}

\begin{table*}[ht!]
\caption{{Mean signatures together with standard deviations for the three clusters of the city performance.} \label{tab:clustering}}
\renewcommand{\tabcolsep}{1pc}\renewcommand{%
\arraystretch}{1.2}%
\vspace{5pt}
\begin{tabular}{c | c | c | c | c | c | c |}
\hline
Cluster & \multicolumn{3}{c|}{LUZs} & \multicolumn{3}{c|}{FUAs}\\
\hline
& RA & CDE & FA & RA & CDE & FA \\
\hline
green  &0.040$\pm$0.038&0.148$\pm$0.054&0.523$\pm$0.284&0.044 $\pm$0.036&0.161$\pm$0.082&0.427$\pm$0.384\\
yellow &-0.008$\pm$0.026&0.032$\pm$0.127&-0.147$\pm$0.133&-0.033$\pm$0.018&0.068$\pm$0.091&-0.088$\pm$0.337\\
orange &-0.042$\pm$0.017&-0.364$\pm$0.160&-0.440$\pm$0.073&-0.001$\pm$0.028& -0.181$\pm$0.114&-0.231$\pm$0.233\\
\end{tabular}
\end{table*}

\begin{table*}[ht!]
\caption{{Similarity measures between the different variations of LUZ and FUA clusterings presented in this paper}   (keeping only FUAs that correspond to any of LUZ cities).\label{tab:similarities}}
\renewcommand{\tabcolsep}{1pc}\renewcommand{%
\arraystretch}{1.2}%
\vspace{2pt}
\begin{tabular}{c | rr}
Clusters  &  $\mathcal{R}$ ( $\bar{\mathcal{R}_\mathrm{r}}$) & $N_{flips}$\\ \hline 
Transactions-based city performance clusters (k=2) & 0.913(0.503) & 1 \\
Transactions-based city performance clusters (k=3) & 0.704(0.538) & 6\\
Socioeconomic clusters (k=2)  & 1.000(0.501)  & 0 \\ 
Socioeconomic clusters (k=3) & 0.771(0.537)  & 3 \\
\hline
\end{tabular}
\\[5pt]
$\bar{\mathcal{R}_\mathrm{r}}$ give the baseline for Rand's criterion $\mathcal{R}$. The closer $\mathcal{R}$ to 1, the better the overlap of between LUZ and FUA clusters. The~baseline values are computed over 10000 random shuffling of the attribution of FUA's clusters. $N_{flips}$ is the minimum number of FUAs to flip from one cluster to another to get a perfect match.
\end{table*}

To investigate our observations in the quantitative manner, we considered the set of three residual metrics as the distinctive signature for each city, and performed a clustering based on signatures similarity using the standard k-means algorithm \cite{macqueen1967kmeans}. Specifically, signatures were defined with the z-score of residuals, which helped to bring different residuals into the common scale and equalized their impact regardless of the original scaling trend variability. Experimenting with the number of clusters we observed a consistent hierarchy between the two- and three-categories cases [figure \ref{fig::clustering3}]. In both cases of LUZs and FUAs the bi-partitioning first revealed the principle distinction between the green cluster and the rest of the country. Furthermore, the clustering in two categories was perfectly consistent for both city definitions, which may serve as an additional evidence of the strength of revealed pattern. A division into three clusters kept the green cluster unchanged  (with the only exception of Granada for FUAs), while the other one split into the yellow and orange clusters. Especially in case of FUAs this split introduced a meaningful separation into borderline cities and the inner part of the country. Higher numbers of clusters did not introduce any new significant patterns, and did not improve the clustering performance in terms of standard metric of goodness such as silhouette. Therefore, for the further analysis we stick to the optimal and the most meaningful clustering into three city categories.

The mean signatures of the clusters together with their internal deviations can be read from table \ref{tab:clustering}. The interpretation is consistent with the previously described city profiles. In both cases, of LUZs and FUAs, the green cluster groups cities that over-perform in all three metrics, which means higher than expected activity of their residents, more transactions executed in a city, as well as higher attractiveness for the incoming tourists, with only few minor exceptions. The orange cluster includes cities that on average perform under the expectations indicated by the scaling trends. At the level of LUZs, they are characterized by the negative values for all three metrics, while for FUAs, apart from clearly negative cases, several cities score moderately within RA and FA parameters. The third cluster -  the yellow one - groups mixed profiles. On average it points to the under-performance in RA and FA with a slight over-performance in CDE [table \ref{tab:clustering}], however, values of particular cities indicate bigger diversity [figure \ref{fig::clustering3}]. 

Received clusters can be further interpreted from the perspective of their spatial organization or overall attractiveness. The green cluster of the well performing cities groups popular touristic destinations, attractive either in terms of a natural environment and leisure potential (cities along the northern coast and islands as well as Barcelona) or in terms of their cultural heritage (e.g. UNESCO site in Toledo, Santiago de Compostela). The case of Toledo is especially interesting as, being the smallest cities in our set, it performs surprisingly well within all three parameters, scoring second in terms of FUA and third for LUZ total performance ranking [see figure \ref{fig::residuals}]. Absence of the country capital among the highly performing cities is somehow surprising, but can be most probably attributed to the complex nature, high inhomogeneity and multiple functions of the capital city, as well as significant size affecting expected performance score. The orange cluster of under-performing cities is especially meaningful with the increased set of cities at the level of FUAs. It covers areas along the northern and southern coastline of Spain, with the exclusion of the most popular destinations attached to the green cluster. To certain extent it can be correlated with a decreased accessibility of particular regions, which may in turn result in a decreased inflow of visitors (e.g. low FA scores), as well as a hindered outflow of residents, decreasing their opportunities for spending diversification.  We can assume decreased accessibility for the cities located in the remote areas such as Galicia, the region of Santiago de Compostela, Vigo and La Coru\'na, Basque Country, but also in the southern parts of Spain (Murcia, Cartagena, or C\'adiz), which are not connected to the high-speed train network. On the contrary, all main capitals Madrid, Seville, Zaragoza, Valladolid, Barcelona, Valencia and M\'alaga, were in 2011 connected with the first range high capacity transport infrastructures. All of them belong either to the green or to the yellow cluster.

To quantify the match between the LUZ and FUA clustering (keeping only the FUA that correspond to a LUZ), we applied the classical Rand's criterion $\mathcal{R}$ \cite{rand1971}. Given two partitions of a set of objects, the Rand criterion measures the fraction of pairs of objects either in the same cluster in both partitions or in different cluster in both partitions. We also propose a simple intuitive way to quantify the match between LUZ and FUA, as the number of FUA to flip from one cluster to another to get a perfect match.   
The LUZ and FUA clusters presented in figure \ref{fig::clustering3} are significantly similar, with a Rand index $\mathcal{R}=0.704$ compared to a baseline index $\bar{\mathcal{R}_\mathrm{r}}=0.538$ [table \ref{tab:similarities}]. One would still need to flip $6$ FUA to another cluster to get a perfect match. At the same time for $k=2$ the match is nearly perfect.

\begin{figure*}[t!]
\centering
\includegraphics[width=0.8\linewidth]{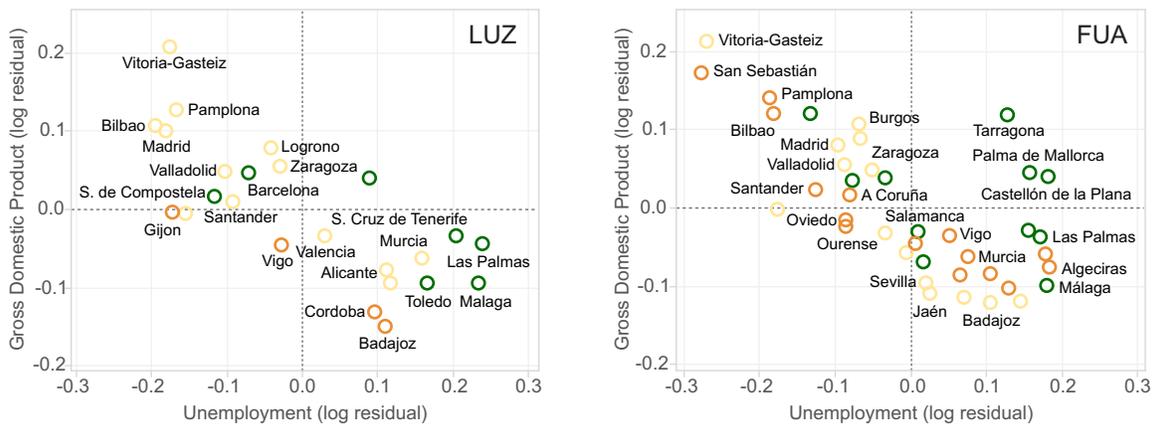}
\caption{\label{fig::economic_distribution}Distribution of socioeconomic urban statistics within the three categories of urban performance (colors coherent with the clusters presented in figure \ref{fig::clustering3}). Gross Domestic Product and unemployment are quantified as log residuals from their respective scaling trends.}
\end{figure*}

\begin{figure*}[ht!]
\centering
\vspace{-5pt}
\includegraphics[width=0.43\linewidth]{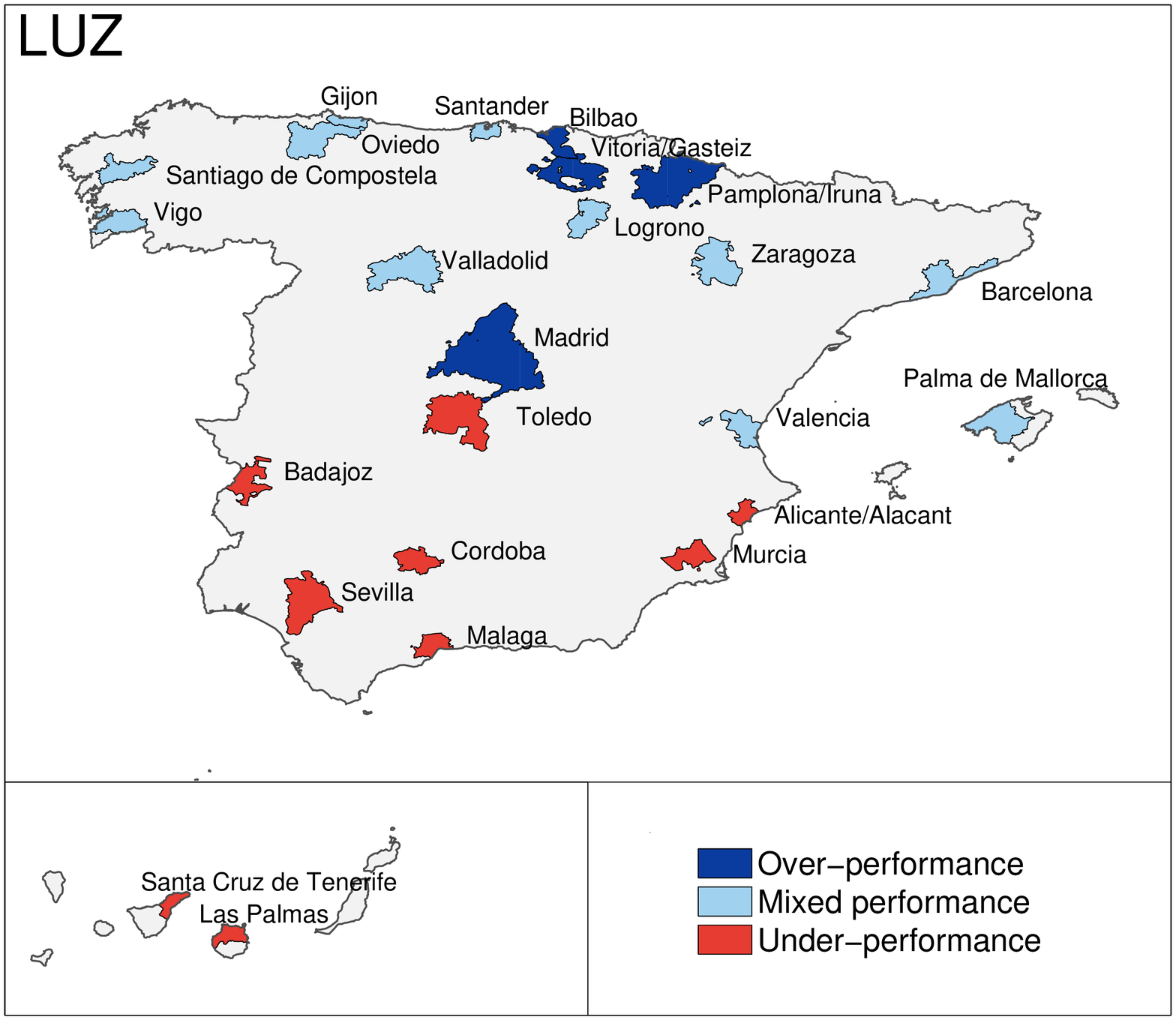}
\includegraphics[width=0.43\linewidth]{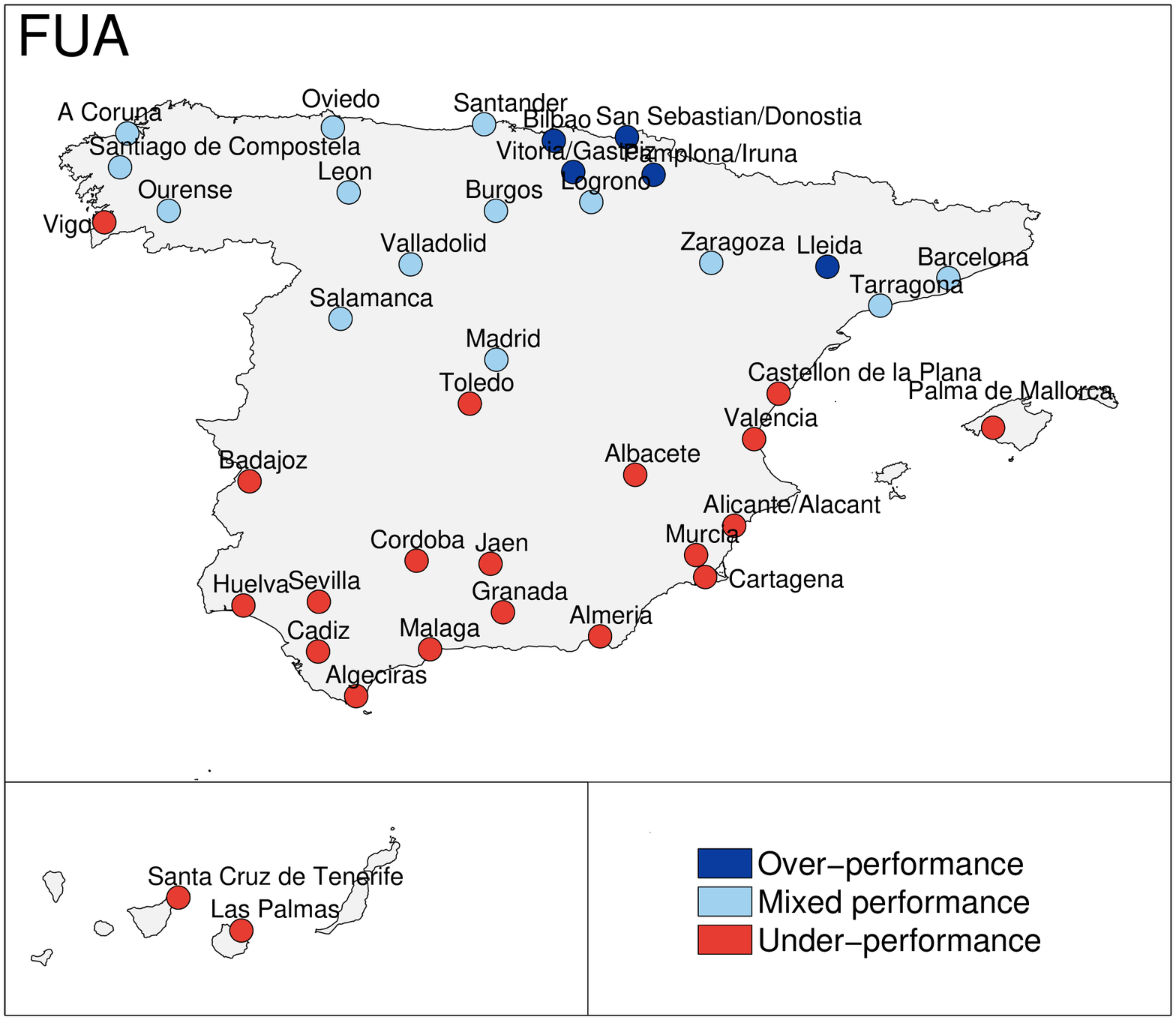}
\vspace{-20pt}
\caption{\label{fig::economic_clustering3}Different categories of cities based on the clustering of socioeconomic urban statistics (Gross Domestic Product and unemployment, quantified as log residuals from the respective scaling trends) into k=2,3 clusters. The partitioning for k=2 can be obtained by considering dark and light blue clusters (over- and mixed performance) as one.}
\end{figure*}

\section{Socioeconomic diversity of \newline{detected clusters}}

In the next step, we explored if and how received clusters of city performance correlate with the standard socioeconomic statistics. As the major economic indicator we considered Gross Domestic Product (GDP), while the social dimension was represented by an unemployment. Both of those parameters were proven to scale superlinearily with city size  \cite {bettencourt2007growth}.Therefore, and for the sake of consistency, we constructed socioeconomic metrics in the similar manner as our performance metrics, i.e. based on the log deviations from the scaling estimates. Distribution of the socioeconomic metrics within the three clusters is presented in figure \ref{fig::economic_distribution}. No clear trend nor correlation can be observed. Instead, we see that each of the clusters, both for LUZs and FUAs, includes cities which are very diverse in terms of  the socioeconomic indicators. The only pattern that can be distinguish is generally high unemployment scores for the cities grouped within the green cluster. Therefore we may suppose that over-performing of those cities, described in previous section, comes also with a price of social problems. 

While clustered, socioeconomic statistics reveal a different spatial pattern compared to the one received with our urban performance metrics. In this case we observe north-south division of Spain [figure \ref{fig::economic_clustering3}], which remains perfectly consistent for both city levels [see table \ref{tab:similarities}]. This pattern is well-know as the major social diversification of the country, emerging not only from an income or employment, but from the vast number of other parameters such as mortality or education indicators measured by the National Statistics Institute of Spain \cite{ine}. General lack of correspondence between the two approaches indicates that the character of clusters detected by means of bank card transactions is principally different than the one captured with the standard socioeconomic statistics. Instead, the novel dataset provide a complementary insights on city performance, coming from the direct observations of the individual economic behavior.

\section{Spending patterns as social \newline{descriptors}}

\begin{figure*}[t!]
\centering
\includegraphics[width=0.99\linewidth]{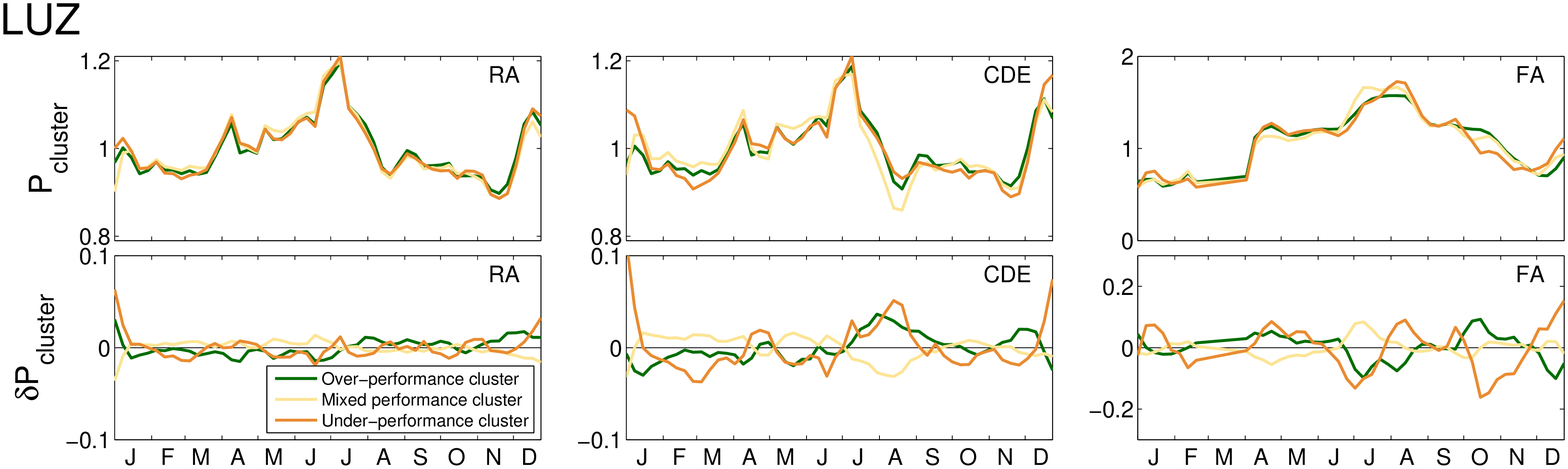}
\includegraphics[width=0.99\linewidth]{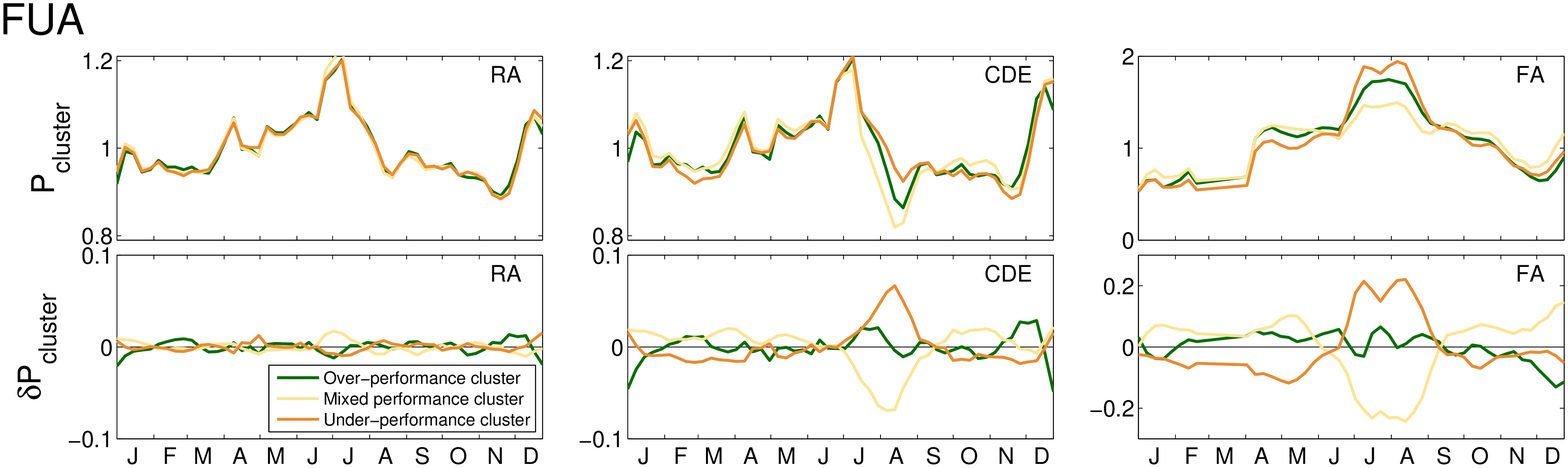}
\caption{ \label{clustersTLW}  Yearly patterns $P$ and deviations $\delta P$ (from the reference case) of the detected clusters at both LUZ and FUA scales. }
\end{figure*}

\begin{figure*}[t!]
\centering

\includegraphics[width=0.45\linewidth]{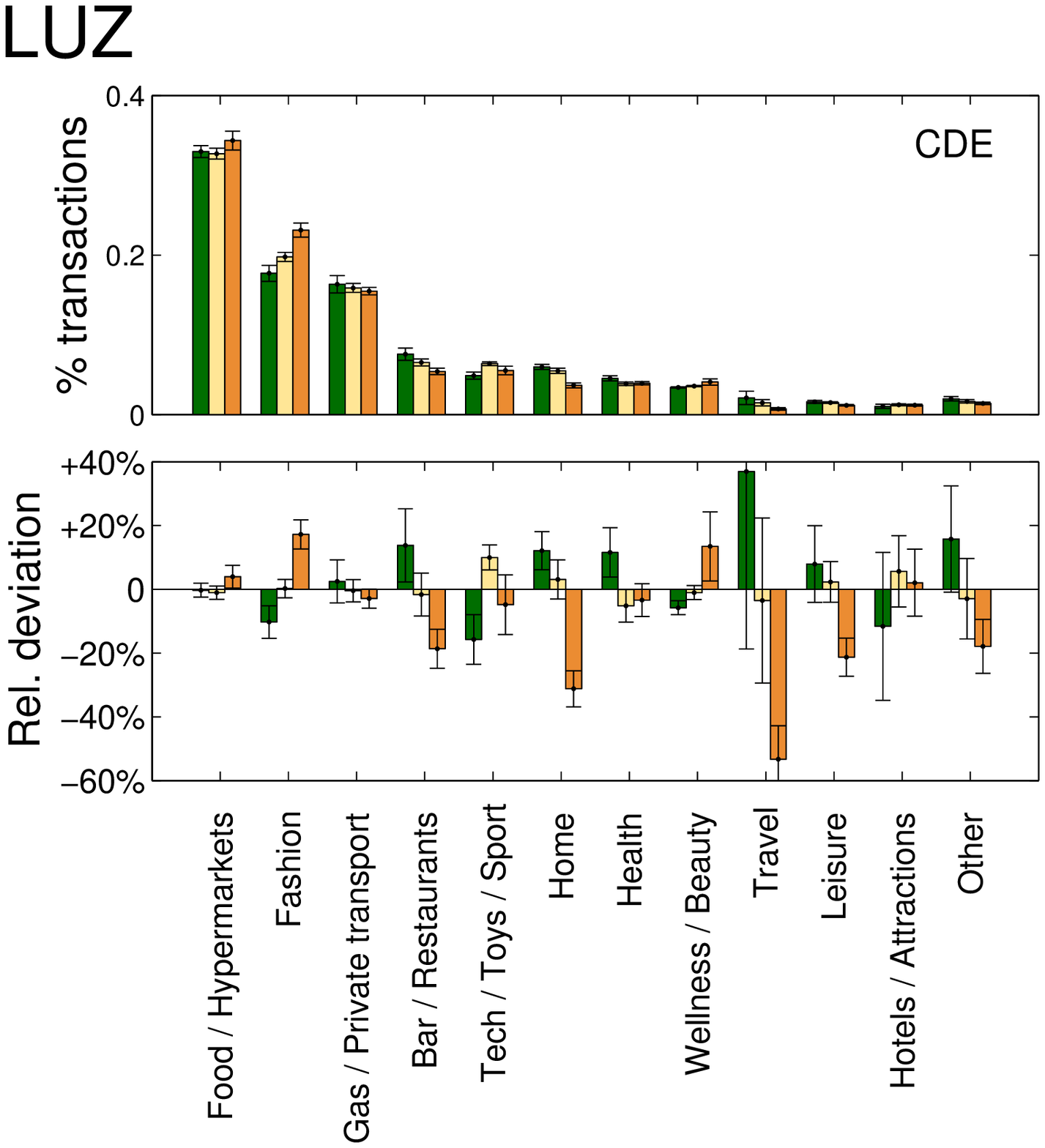}
\includegraphics[width=0.45\linewidth]{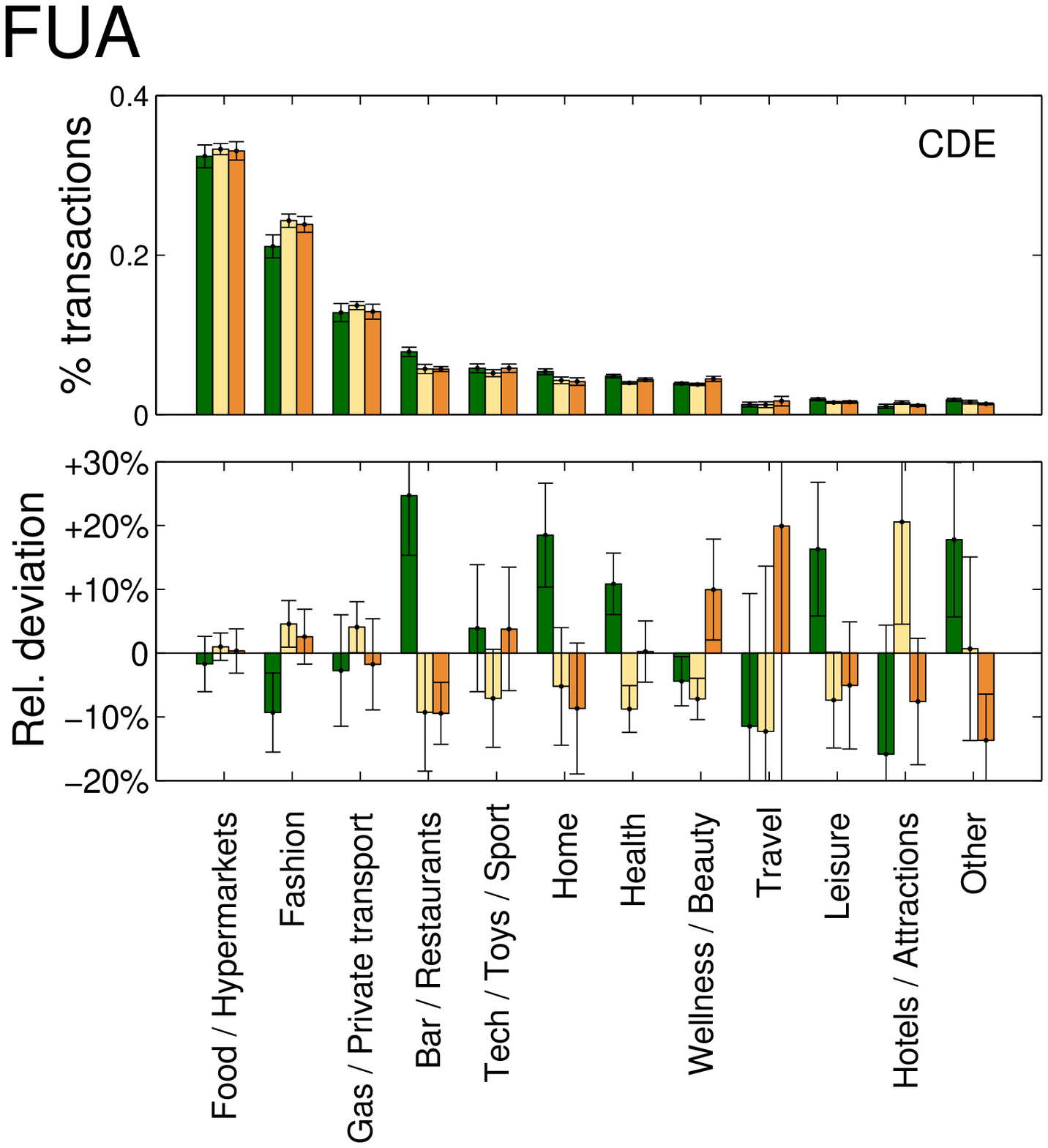}
\vspace{-5pt}
\caption{\label{fig::clustersCATprofile} Average spending profiles of cities grouped according to three urban performance clusters (top) and corresponding relative deviations from the cities' average profile (bottom). Spending profile records percentages of transactions performed in the 12 major business categories. Error bars indicate standard deviations from the mean of each cluster.}
\end{figure*}

In addition to the three major city metrics (RA, CDE and FA residuals) analyzed before, we explored the transactions data from two other angles to better characterize the detected clusters by spending behavior of people within them.

First, we investigated temporal fluctuations of people's spending. To this end, we characterized each city by its {\it yearly pattern}, defined as the mean-normalized, weekly aggregated number of transactions being made in that city during the whole year - the normalization allowing us to compare the relative fluctuations of activities between cities. Figure \ref{clustersTLW} shows the average yearly pattern of the cities within each cluster for the three different groups of individuals at both LUZ and FUA scales. 
All three clusters share roughly similar patterns.
RA and CDE patterns don't fluctuate much during the year~- the~weekly spending varying from $80\%$ to $120\%$ of its average value - with localized peaks of activity on Easter, in early July (which may correspond to sales) and late December (for the Christmas and New Year holidays). FA patterns show higher fluctuations, the activity gradually going from $50\%$ - during the winter - to around $200\%$ of the year average in the summer, when most tourists are visiting Spain.

The singularity of each clusters can be readily appreciated when looking at the patterns' deviations from the natural reference case built by averaging the patterns of all cities [figure \ref{clustersTLW}]. 
RA deviation patterns give us insights on the spending behavior of individuals: at both LUZ and FUA scales, they are all always close to zero, meaning that the average number of transactions made by individuals at a given period of the year does not depend on their city of origins. On the other hand, CDE and FA deviations patterns illustrate how the economic activity within cities differ from one cluster to another. They are particularly interesting at the FUA scale, where they show how the summer period impacts in different ways the three clusters. Cities from the orange cluster (along the Mediterranean and Atlantic coasts) show positive deviation from the average cities (of about $5\%$ for CDE and $20 \%$ for FA), revealing a strong specialization of these cities in tourism. This over-specialization reflect the existence of a large proportion of part-year employment, recognized in particular as a main problem on the Mediterranean coast \cite{del2007plan}, which can explain their under-performance in terms of our residuals analysis. On the contrary, cities from the yellow cluster - such as Madrid - show a negative deviation from the average cities (of about $5\%$ for CDE and $20 \%$ for FA), revealing a relative lower focus on touristic sector that may partly explain their middle-range performance. Finally, cities from the green cluster don't show a significant deviation from the normal impact of summer, which may reflect a well balanced development between the touristic and other economic sectors and explain their high performance.
Similar interpretations can not readily be made for the LUZ clusters, albeit the case of CDE deviation patterns also show an impact of summer activities. 

Next, we examined the clusters in terms of people's spending preferences. To this end, we defined for each city a {\it spending profile} recording the percentage of transactions made by domestic individuals in the 12 major categories of businesses.
\footnote{We did not perform a similar analysis for foreign customers' spending profiles as those might be highly biased by the inhomogeneity of BBVA card terminals share among businesses of different categories.}
 Profile of a cluster is defined as the average profile of the cities within this cluster [figure \ref{fig::clustersCATprofile}].
At a glance, the profiles of different clusters are roughly similar for the basic households expenses such as food, gas or clothes (around $35\%$ of the budget is devoted to the Food/Hypermarket category, around $20\%$ for Gas/Private Transport and around $15\%$ for Fashion, with some variability depending on the city definition level). 

Unique characteristic of each cluster appears more clearly when analyzing the relative deviation from the average of all cities spending profiles [figure \ref{fig::clustersCATprofile}]. Quite remarkably, we observe a good match between LUZ and FUA results. The relative deviation of the percentage of transactions made for basic items such as food, fashion and gas in the different clusters are not significant (as can be seen from the error bars which overlap the x-axis), although green cluster appear to have slightly less transactions, and yellow and orange clusters slightly more transactions than the average in this categories. This is especially visible for the Fashion category, where the green cluster possesses around $10\%$ less transactions than the average. Other more significant deviations from the average pattern for cities from the green cluster include positive deviations in Bars and Restaurants (+$35\%$ than the average city in FUAs), Home appliance stores (+$18\%$ in FUAs), Health (+$11\%$ in FUAs), Leisure (+$17\%$ in FUAs) and in the `Other' category, which combines different small businesses and services. On the contrary, cities from the yellow and orange clusters make relatively less transactions than the average in all those categories. 

\newpage
To summarize, in over-performing cities, the weight of primary necessity items on the family budget appears generally lower than in mixed and under-performing cities, which is combined with a larger amount spend in categories that tend to increase the quality of life. Could their over-performance be explained by a larger availability of goods and services increasing individuals' well-being?   

\section{Conclusions}

In this paper we confirmed the applicability of the new source of large-scale individual economic data, i.e. countrywide records of bank card transactions, to characterize and compare urban performance independently of city scale. We proposed to describe cities with three parameters related to the economic activity of city residents and foreign visitors as well as earnings of city businesses. We found those parameters to scale superlinerily with city size, which well agrees with the previous studies of urban features. 
Based on this observation, we were able to address urban performance in a scale-independent manner - through the deviations from respective scaling laws showing particular city over- or under- performance compared to the general trend. 
We further demonstrated that city comparison and automated classification grounded in the proposed metrics lead to a~discovery of the meaningful spatial and temporal patterns. Suggested methodology was employed based on the example of Spain, but being of a generic character it would be easily applicable for the assessment of cities in other countries or regions.  

The analysis revealed important insights on the characteristics of Spanish cities, which significantly complement the patterns that might be discovered based on the standard socioeconomic statistics, and which are of a potential interest for urban authorities and planners. Automated clustering of the proposed performance metrics indicated three profiles of cities, that were proven to further correlate with the specific spending behaviors. The most significant and clearly distinguished group included the major cultural and natural touristic destinations of the country. Those cities were consequently over-performing in all metrics and also indicated increased tendency for certain purchases e.g. connected to leisure and higher quality of life. On the contrary, cities located along the other parts of the Spanish coast revealed significant under-performance in terms of all types of transactional activity. This finding can be associated with their over-specialization on summer tourism proved by the specific temporal pattern of visitors' spending.  Importantly, the majority of observed patterns were surprisingly consistent for the two tested levels of a city definition, pointing to the robustness of the proposed methodology regardless of a spatial scale of the analysis.  

\section{Acknowledgement}

The authors would like to thank the Banco Bilbao Vizcaya Argentaria (BBVA) for providing the dataset for this research. Special thanks to Assaf Biderman, Marco Bressan, Elena Alfaro Martinez and Mar\'ia Hern\'andez Rubio for organizational support of the project and stimulating discussions. We further thank the BBVA, Ericsson, MIT SMART Program, Center for Complex Engineering Systems (CCES) at KACST and MIT, the National Science Foundation,the MIT Portugal Program, the AT\&T Foundation, Audi Volkswagen, The Coca Cola Company, Expo 2015, Ferrovial, The Regional Municipality of Wood Buffalo and all the members of the MIT Senseable City Lab Consortium for supporting the research. This research was also partially supported by the Austrian Science Fund (FWF) through the Doctoral College GIScience (DK W 1237-N23).

\end{document}